\documentclass[a4paper]{jpconf}
\usepackage{graphicx}
\usepackage{mathtools}
\begin{document}

\title{$K \rightarrow \pi \ell^+ \ell^-$: Status and update\footnote{Talk given at the KAON 2016 Conference, 
Birmingham (UK) (14-17 September, 2016). }}

\author{Jorge Portol\'es}

\address{IFIC, Universitat de Val\`encia -- CSIC, Apt. Correus 22085, E-46071 Val\`encia, Spain}

\ead{Jorge.Portoles@ific.uv.es}

\begin{abstract}
$K \rightarrow \pi \ell^+ \ell^-$ decays, $\ell=e, \mu$,  will be one of the goals of future kaon facilities, like NA62 (CERN) and K$^0$TO 
(J-PARC). 
We review briefly the theoretical status of those processes. 
\end{abstract}

\section{Introduction}

Suppressed (rare) kaon decays in the Standard Model (SM) constitute an interesting benchmark where to look for New Physics. Common sense and logic tell us that, in any observable, in order to look for physics beyond the SM one needs to have a reliable SM prediction and this is not obvious in many kaon decays. Indeed low-energy strong interacting effects, of difficult implementation, may be involved in 
those determinations \cite{Cirigliano:2011ny}.  
\par 
$K \rightarrow \pi \nu \overline{\nu}$ are atypical, poorly measured, decays that are dominated by short-distance interactions 
\cite{Buchalla:1993wq,Buras:2005gr} and, accordingly, have a very accurate prediction within the SM. This is the reason why they are the main goal of present dedicated facilities: $K^+ \rightarrow \pi^+ \nu \overline{\nu}$ at NA62 (CERN) and $K_L \rightarrow \pi^0 \nu \overline{\nu}$
at K$^0$TO (J-PARC) \cite{Komatsubara:2012pn}. These devices should be able, in addition, to get samples of other related processes 
like $K^+ \rightarrow \pi^+ \ell^+ \ell^-$ ($\ell = e, \mu$) at NA62 or $K_L \rightarrow \pi^0 \gamma \gamma$ at K$^0$TO (as we will comment the latter is relevant to determine the CP-conserving contribution to $K_L \rightarrow \pi^0 \ell^+ \ell^-$).
\par 
In this note we summarize the theoretical status of the $K \rightarrow \pi \ell^+ \ell^-$ decays. These are  $\Delta S = 1$
weak neutral current transitions in the SM and, hence, very suppressed . Contrarily to
the case above, these decays are fairly dominated by strongly interacting low-energy dynamics which is always harder to determine. 
As in many other processes of analogous features, Chiral Perturbation Theory (ChPT) is the appropriate tool to handle these decays
\cite{Gasser:1983yg,Gasser:1984gg}. They were first studied within this framework by Ecker, Pich and De Rafael \cite{Ecker:1987qi,Ecker:1987hd} when they computed the dominant octet ($\Delta I=1/2$) contributions at the leading ${\cal O}(p^4)$ in ChPT. This set up a systematic approach that would pave the way for later improvements, as we will see. The subdominant contributions of $SU(3)$ 27-representation operators, at 
${\cal O}(p^4)$, have been carried out in Ref.~\cite{Ananthanarayan:2012hu}. 
\par 
$K^+, K_S \rightarrow \pi \ell^+ \ell^-$ decays are dominated by single virtual-photon exchange, $K \rightarrow \pi \gamma^*$. However this contribution is CP violating for $K_L \rightarrow \pi^0 \ell^+ \ell^-$ and, consequently, the latter is driven by substantially different dynamics. The theoretical status of the first processes will be set forth in Section~1. Meanwhile Section~2 will be dedicated to the 
description of the $K_L$ decay. 
\par 
A more detailed account of $K \rightarrow \pi \ell^+ \ell^-$ decays and their corresponding bibliography can be looked up in Ref.~\cite{Cirigliano:2011ny}.

\section{$K^+, K_S \rightarrow \pi \ell^+ \ell^-$}

CP allowed $K(k) \rightarrow \pi(p) \ell^+ \ell^-$, $\ell=e,\mu$, decays are dominated by single virtual photon exchange
$K \rightarrow \pi \gamma^*$. This is the case for $K^+$ and $K_S$ decays. Hence the amplitude is determined by one electromagnetic transition
form factor in the presence of nonleptonic weak interactions \cite{Ecker:1987qi,D'Ambrosio:1998yj}:
\begin{equation} \label{eq:gf}
i \int d^4x \, e^{iqx}  \langle \pi(p) | \, T \left\{ J_{em}^{\mu}(x) \, {\cal L}_{\Delta S = 1}(0) \right\} | K_j(k) \rangle  = 
\frac{G_F M_K^2}{(4 \pi)^2} \, V_j(z)  \left[ z  (k+p)^{\mu} - (1-r_{\pi}^2)  q^{\mu} \right], 
\end{equation}
with $j=+,S$, to distinguish both decays, and $q = k-p$, $z=q^2/M_K^2$, $r_P = M_P/M_K$. Here $J_{em}^{\mu}$ is the electromagnetic current  and
$ {\cal L}_{\Delta S=1}$ the strangeness changing nonleptonic weak Lagrangian. The spectrum in the dilepton invariant mass is 
then given by:
\begin{equation} \label{eq:sp}
\frac{d \, \Gamma_j}{d \, z} \, = \, \frac{G_F^2 \alpha^2 M_K^5}{12 \pi (4 \pi)^4} \, \lambda^{3/2}(1,z,r_{\pi}^2) \,  \sqrt{1-4 \frac{r_{\ell}^2}{z}} \,  \left(1+2 \frac{r_{\ell}^2}{z}\right) \, |V_j(z)|^2 \, ,
\end{equation}
with $\lambda(a,b,c)=(a+b-c)^2-4ab$ and $4 r_{\ell}^2 \leq z \leq (1-r_{\pi})^2$.
\par
\begin{figure}[h]
\begin{center}
%\begin{minipage}{18pc}
\includegraphics[width=13pc]{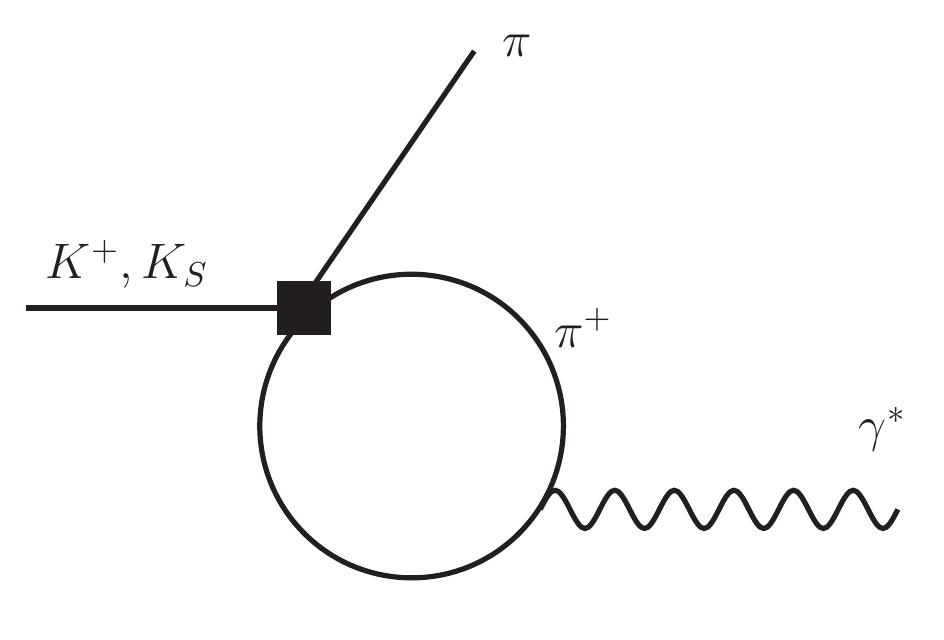}
\caption{\label{fig:01} $K \rightarrow \pi \pi \pi$ contribution to $K^+, K_S \rightarrow \pi \gamma^*$.}
%\end{minipage}
\end{center}
\end{figure}

 
The $V_j(z)$ form factor can be determined within the ChPT framework, i.e. in a model-independent way. 
Because gauge invariance, $V_j(z)=0$ at ${\cal O}(p^2)$ in the
chiral expansion. Hence the leading contribution is ${\cal O}(p^4)$ that was computed in Ref.~\cite{Ecker:1987qi}. Although a complete study
to next-to-leading ${\cal O}(p^6)$ has not been performed yet, the dominant unitarity corrections from $K \rightarrow \pi \pi \pi$, 
shown in Fig.~\ref{fig:01}, were studied in Ref.~\cite{D'Ambrosio:1998yj}, that constitute the present {\em state-of-the-art} (there is also
a tiny contribution from a kaon running in the loop, that obviously has no absorptive part).
The full result can be decomposed in a polynomial contribution (linear in $z$ up to this order) plus a unitarity loop correction~:
\begin{equation} \label{eq:vjz}
V_j(z) \, = \, a_j \, + \, b_j \, z \, + \, V_j^{\pi \pi}(z), \; \; \; \; \; \; \qquad (j=+,S).
\end{equation}
The polynomial part is, unfortunately, not fully given by chiral symmetry. It translates, mainly, into the Low Energy Couplings (LECs) of
the ChPT framework, and it encodes contributions from higher energy spectra that have been integrated out in the process to reach the low-energy effective theory. $a_j$ gets contributions starting at ${\cal O}(p^4)$ while $b_j$, but for a tiny ${\cal O}(p^4)$ kaon loop input in Fig.~\ref{fig:01}, starts at ${\cal O}(p^6)$. As we will comment we have, up to now, only model-dependent theoretical predictions for these parameters.
\par 
The unitarity corrections in $V_j(z)$ are given by~:
\begin{equation} \label{eq:uk3p}
V_j^{\pi \pi}(z) \, = \, \overbracket{\frac{\alpha_j + \beta_j (z-z_0)/r_{\pi}^2}{G_F M_K^2 r_{\pi}^2}}^{K \rightarrow \pi \pi \pi} \,
\overbracket{\left[ \frac{4}{9} - \frac{4}{3z} + \frac{4}{3z} \left( 1-\frac{z}{4} \right) \, G(z) \right]}^\text{1-loop} \, \overbracket{\left[ 1 + \frac{z}{r_{\rho}^2} \right]}^{F_V(z)} ,
\end{equation}
where $z_0 = r_{\pi}^2 + 1/3$. We now explain the structure of the unitarity corrections:
\begin{itemize}
\item The interaction of the vertex $K \rightarrow \pi \pi \pi$ in Fig.~\ref{fig:01} is driven by the $\alpha_j$ and $\beta_j$ constants that
are expressed in terms of the parameterizations of Ref.~\cite{Kambor:1991ah}:
\begin{eqnarray} \label{eq:dd}
\alpha_+ &=& \beta_1 - \frac{1}{2} \beta_3 + \sqrt{3} \gamma_3 \, , \qquad \beta_+ \, = \, 2( \xi_1 + \xi_3 - \xi_3') \, \nonumber \\
\alpha_S &=&  - \frac{4}{\sqrt{3}} \gamma_3 \, , \qquad \qquad \qquad \! \! \beta_S = \frac{8}{3} \xi_3'. 
\end{eqnarray}
It can be noticed that $\alpha_+, \beta_+$ are given by both $\Delta I=1/2, 3/2$ transitions, while $\alpha_S, \beta_S$ come only from $\Delta I=3/2$ 
transitions, because these are the only ones that drive the CP conserving contribution to $K_S \rightarrow \pi^0 \pi^+ \pi^-$.  
As a consequence $V_S^{\pi \pi}(z)$ is fairly suppressed in comparison with $V_+^{\pi \pi}(z)$. 
\item The 1-loop function $G(z)$ is given in Ref.~\cite{Cirigliano:2011ny}.
%\begin{equation} \label{eq:gz}
%G(z) = \left\{ \begin{array}{lr}
%                 \sqrt{4/z-1} \arcsin \left( \frac{\sqrt{z}}{2}\right)
%                 & z  \leq 4 \\
%                  \frac{1}{2} \sqrt{1-4/z} \left( \ln
%                 \frac{1+\sqrt{1-4/z}}{1-\sqrt{1-4/z}} - i \pi \right)
%                  & z > 4
%                \end{array}
%        \right. .
%\end{equation}
\item The $\pi^+ \pi^- \gamma^*$ vertex in the one-loop diagram is dressed through the vector form factor in the Resonance Chiral Theory
framework \cite{Ecker:1989yg}:
\begin{equation} \label{eq:fvz}
F_V(z) \, = \, \frac{M_{\rho}^2}{M_{\rho}^2-M_K^2 z} \, \simeq \, 1 + \frac{z}{r_{\rho}^2} \, , 
\end{equation}
\end{itemize}
If, as expected, $a_j, b_j \sim {\cal O}(1)$, the polynomial contribution fairly dominates over the unitarity-cut loop corrections
given by $V_j^{\pi \pi}(z)$. Hence these decays are very sensitive to the chiral LECs. Several models have been considered for their computation. One of them comes to mind immediately from the comparison of the electromagnetic form factor in Eq.~(\ref{eq:fvz}) and the weak polynomial term in $V_j(z)$ (\ref{eq:vjz}). One may consider that this later term could also be dominated by a resonance transition (VMD). This would predict:
\begin{equation} \label{eq:vmd}
\left\{
\; \; \; a_j \, b_j > 0 , \qquad \qquad \frac{b_j}{a_j} \, = \, \frac{M_K^2}{M_{\rho}^2}  \, \simeq \, 0.4 \; \; \; 
\right\}^{\; \mathrm{VMD \; model}} \! \! \! \! \! \! \! \! \! \! \! \! \! \! \! \! \! \! \! \! \! \! \! \! \! .
\end{equation}
Other models have also been employed: weak deformation model \cite{Ecker:1990in}, minimal hadronic approximation \cite{Friot:2004yr},
Pad\'e type \cite{Dubnickova:2006mk} or large-$N_C$ considerations \cite{Coluccio-Leskow:2016tsp}. Recently RCB and UKQCD collaborations 
\cite{Christ:2015aha,Christ:2016mmq} have studied these decays within lattice QCD and have produced:
\begin{equation} \label{eq:a+b+lat}
a_+ = 1.4 \pm 0.7, \qquad \qquad b_+ = 0.7 \pm 0.8 \, ,
\end{equation} 
that are in good agreement with the VMD model prediction (\ref{eq:vmd}). It has to be noted that this is a first try in the lattice.
They are working with unphysically heavy pion and kaon masses and, accordingly, an extrapolation to the physical point is needed. We should
expect an improvement from the lattice in the near future. 
\par 
The values of $a_j, b_j$, can be extracted from the experimental data by using our model-independent description of $V_j(z)$ in Eq.~(\ref{eq:vjz}) as they are the only unknowns. The present situation, given by NA48/1 and NA48/2 is depicted in Table~\ref{tab:1}.
\begin{table} 
\centering
\begin{tabular}{|ccccc|}
\hline
Process & BR $\times 10^9$ & $a_j$ & $b_j$ & $b_j / a_j$ \\
\hline
$K^+ \rightarrow \pi^+ e^+ e^-$ \cite{Batley:2009pv} & $314 \pm 10$ & $-0.578 \pm 0.016$ & $-0.779 \pm 0.066$ & $\sim 1.35$ \\ 
\hline
$K^+ \rightarrow \pi^+ \mu^+ \mu^-$ \cite{Batley:2011zz} & $96.2 \pm 2.5$ & $-0.575 \pm 0.039$ & $-0.813 \pm 0.145$ & $\sim 1.41$ \\
\hline
$K_S \rightarrow \pi^0 e^+ e^-$ \cite{Batley:2003mu} &  $5.8^{+2.9}_{-2.4}$ & $|1.06|^{+0.26}_{-0.21}$ & - & - \\
\hline 
$K_S \rightarrow \pi^0 \mu^+ \mu^-$ \cite{Batley:2004wg} & $2.0^{+1.5}_{-1.2}$ & $|1.54|^{+0.40}_{-0.32}$ & - & - \\
\hline
\end{tabular}
\caption{\label{tab:1} Experimental determinations of $K \rightarrow \pi \ell^+ \ell^-$ by NA48/1 ($K_S$) and NA48/2 ($K^+$).}
\end{table}
While in $K^+$ decay both, branching and spectrum, have been determined, in the $K_S$ decay we only have the branching ratio. 
As a result only the $a_S$ parameter has been resolved in the neutral decay, while we have results for both $a_+$ and $b_+$. 
By comparing the prediction of the VMD model in Eq.~(\ref{eq:vmd}) with the phenomenological results for $a_+$ and $b_+$ in Table~\ref{tab:1} we notice that the only agreement is in the common sign of both parameters but not in their ratio. Hence one could consider either that the model is not correct or that higher order chiral corrections are not negligible. This later hypothesis would be rather unexpected. 
\par
\begin{figure}[h]
\begin{center}
\begin{minipage}{16pc}
\includegraphics[width=15pc]{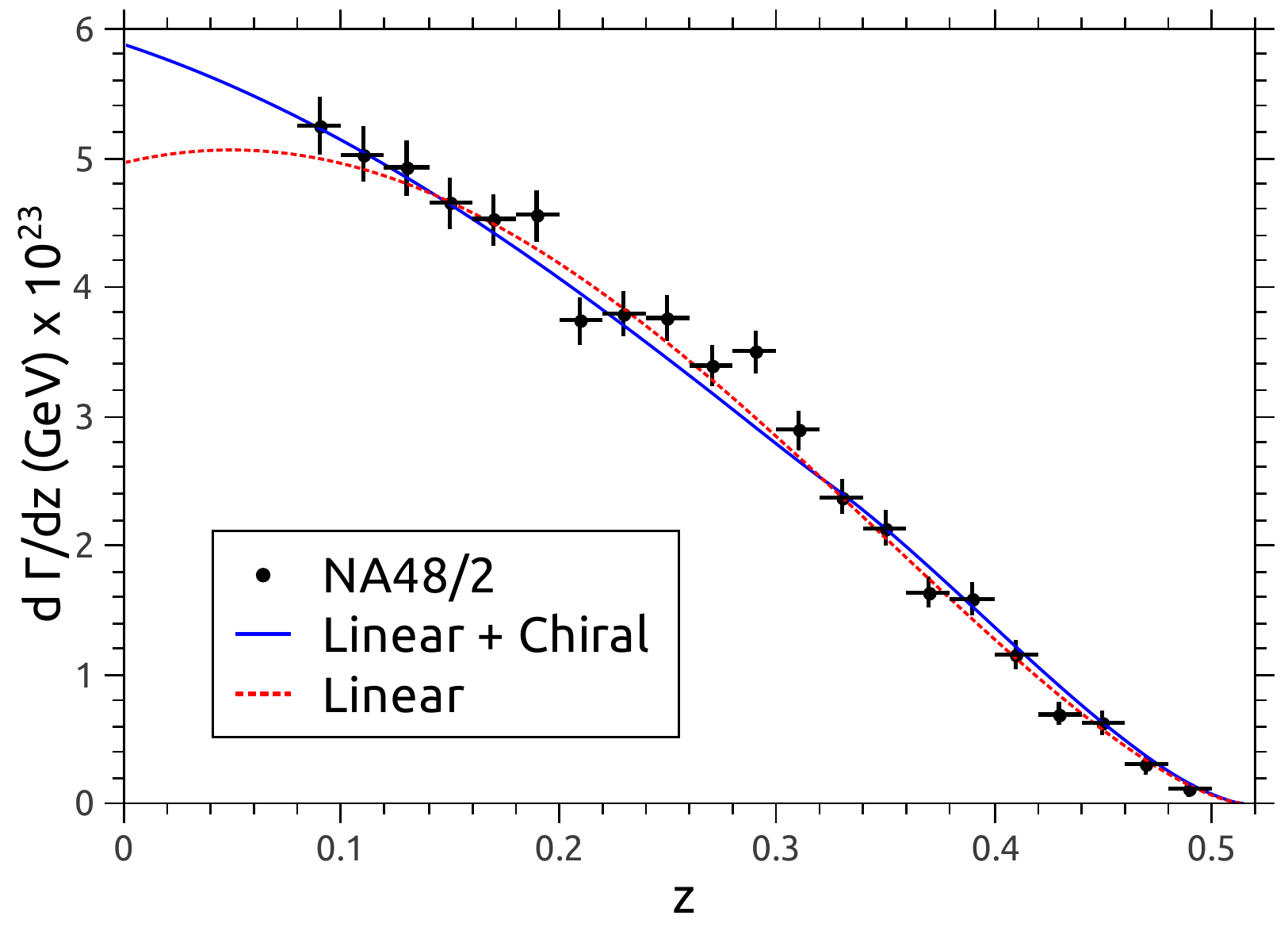}
\caption{\label{fig:02}Spectra for $K^+ \rightarrow \pi^+ e^+ e^-$ in the dilepton invariant mass. NA48/2 data from \cite{Batley:2009pv}.}
\end{minipage}\hspace{2pc}%
\begin{minipage}{16pc}
\includegraphics[width=16pc]{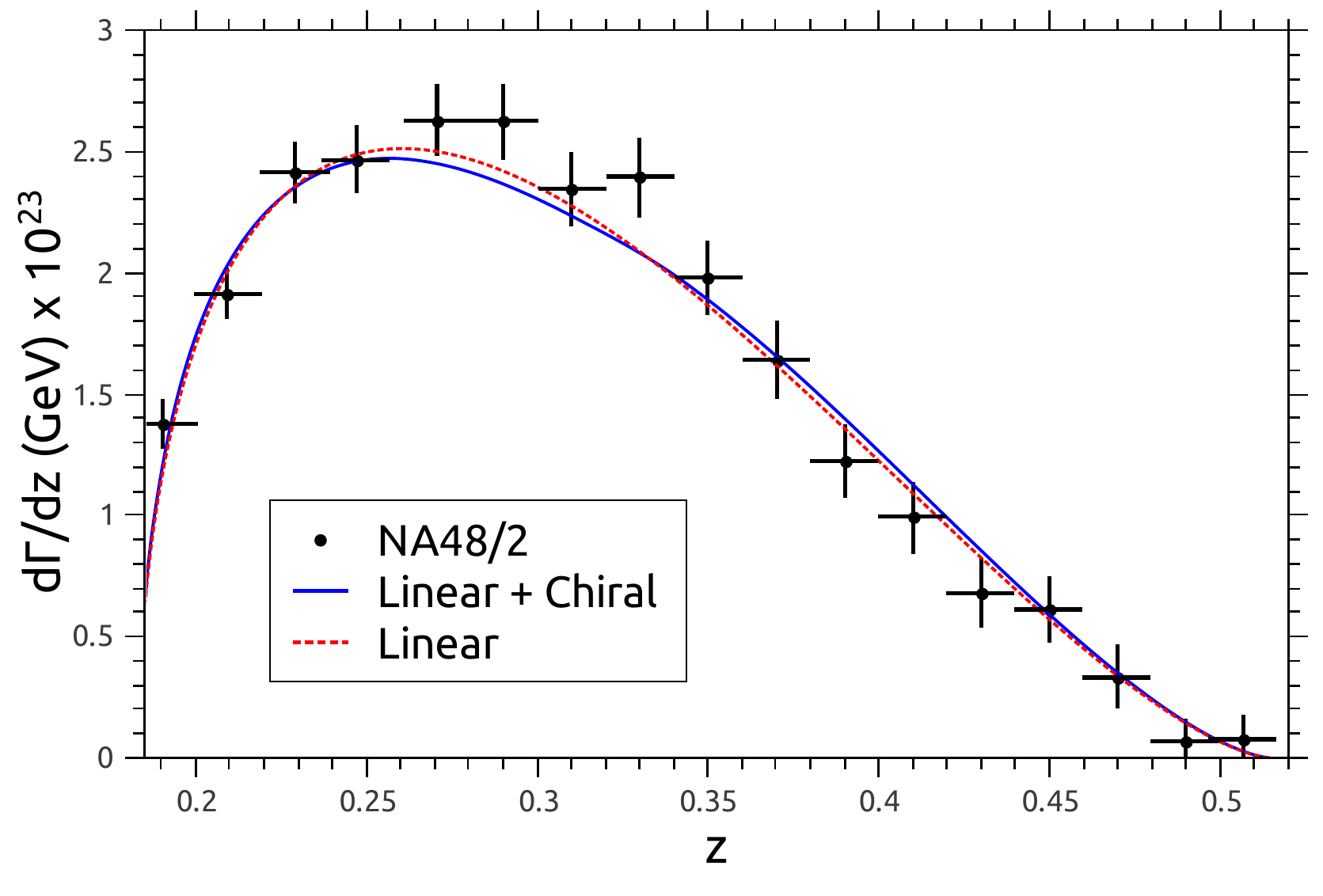}
\caption{\label{fig:03}Spectra for $K^+ \rightarrow \pi^+ \mu^+ \mu^-$ in the dilepton invariant mass. NA48/2 data from \cite{Batley:2011zz}.}
\end{minipage} 
\end{center}
\end{figure}

  
The spectra of $K^+ \rightarrow \pi^+ e^+ e^-$ decay is shown in Figure~\ref{fig:02} while the one of 
$K^+ \rightarrow \pi^+ \mu^+ \mu^-$ is shown in Figure~\ref{fig:03}. In those figures we also show the theoretical prediction
as given by the full $V_+(z)$ in Eq.~(\ref{eq:vjz}) (Linear + Chiral) and keeping the polynomial part only (Linear) for the
$a_+$ and $b_+$ parameters in Table~\ref{tab:1}. As can be
noticed this latter contribution dominates and the role of the unitarity loop is rather small. 
\par 
In addition to the dominant $K^{\pm} \rightarrow \pi^{\pm} \gamma^*$ amplitude, the form factor $V_+(z)$
receives a short-distance contribution from the
operator $Q_{7V}$ from the effective Lagrangian \cite{Cirigliano:2011ny}.
Although negligible in the branching ratios and spectra, its
interference with the long-distance amplitude
leads to a CP-violating charge asymmetry in $K^{\pm} \rightarrow \pi^{\pm} \ell^+ \ell^-$
\cite{Ecker:1987hd}, defined by:
\begin{equation} \label{eq:kpeeasy}
\Delta \Gamma  \equiv  \frac{\Gamma(K^+ \rightarrow \pi^+ e^+ e^-) - \Gamma(K^- \rightarrow \pi^- e^+ e^-)}{\Gamma(K^+ \rightarrow \pi^+ e^+ e^-) + \Gamma(K^- \rightarrow \pi^- e^+ e^-)} 
 \sim  0.07  \, \mbox{Im} \, \lambda_t ,
\end{equation}
where $\lambda_q = V_{qd} V_{qs}^*$. For the numerical determination we have used that $\mbox{Im} \, b_+ / \mbox{Im} \, a_+ \simeq M_K^2/M_{\rho}^2$ (VMD model). With $\mbox{Im} \, \lambda_t \sim \eta \lambda^5 A^2 \, \sim 10^{-4}$, this asymmetry is tiny within the SM
\cite{D'Ambrosio:1998yj,D'Ambrosio:2002fa}. Experimentally the present bound is given by $\Delta \Gamma < 0.021$ at $90 \%$ C.L. 
\cite{Batley:2009pv}. 
\par
\begin{figure}[t]
\begin{center}
\includegraphics[width=17pc]{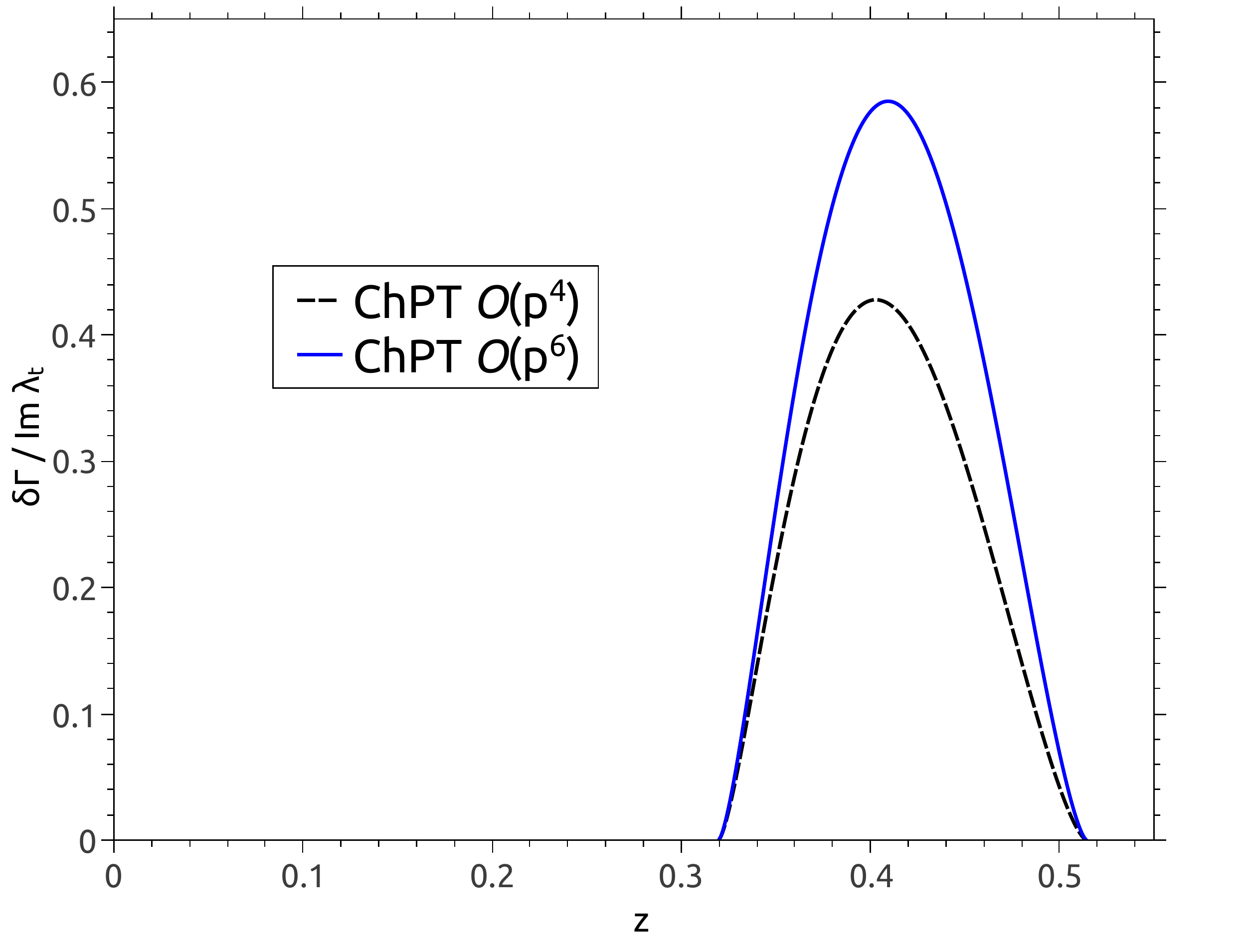}
\caption{\label{fig:04}
ChPT predictions for the unintegrated charge asymmetry for the process $K^+ \rightarrow \pi^+ e^+ e^-$ in the $z$ variable.}
\end{center}
\end{figure}

 
An interesting observable could be the unintegrated charge asymmetry \cite{D'Ambrosio:1998yj} defined by:
\begin{equation} \label{eq:unipeeasy}
\delta \Gamma(z) \equiv \frac{\frac{d \Gamma}{dz} (K^+ \rightarrow \pi^+ e^+ e^-) - \frac{d\Gamma}{dz} (K^- \rightarrow \pi^- e^+ e^-)}{\Gamma(K^+ \rightarrow \pi^+ e^+ e^-) + \Gamma(K^- \rightarrow \pi^- e^+ e^-)} .
\end{equation}
In order to skip the background produced by the Dalitz decay of a $\pi^0$, namely $K^{\pm} \rightarrow \pi^{\pm} \pi^0_D$, 
$\pi^0_D \rightarrow e^+ e^- \gamma$, a cut in the experimental data excludes events with $z < 0.08$. The unintegrated asymmetry, 
as seen in Fig.~\ref{fig:04}, stays a higher values. 
\section{$K_L \rightarrow \pi^0 \ell^+ \ell^-$}

The decay $K_L \rightarrow \pi^0 \gamma^*$ is CP violating. As a consequence the dynamics of the $K_L \rightarrow \pi^0 \ell^+ \ell^-$ decay
is more involved than the processes above. Indeed both long-distance and short-distance contributions drive their observables. The structure of the amplitude includes, in addition to the vector form factor $V_L(z)$, other form factors, namely scalar, pseudoscalar and axial-vector, may contribute to this decay \cite{Isidori:2004rb}. 
\par
\begin{figure}[h]
\begin{center}
%\begin{minipage}{18pc}
\includegraphics[width=14pc]{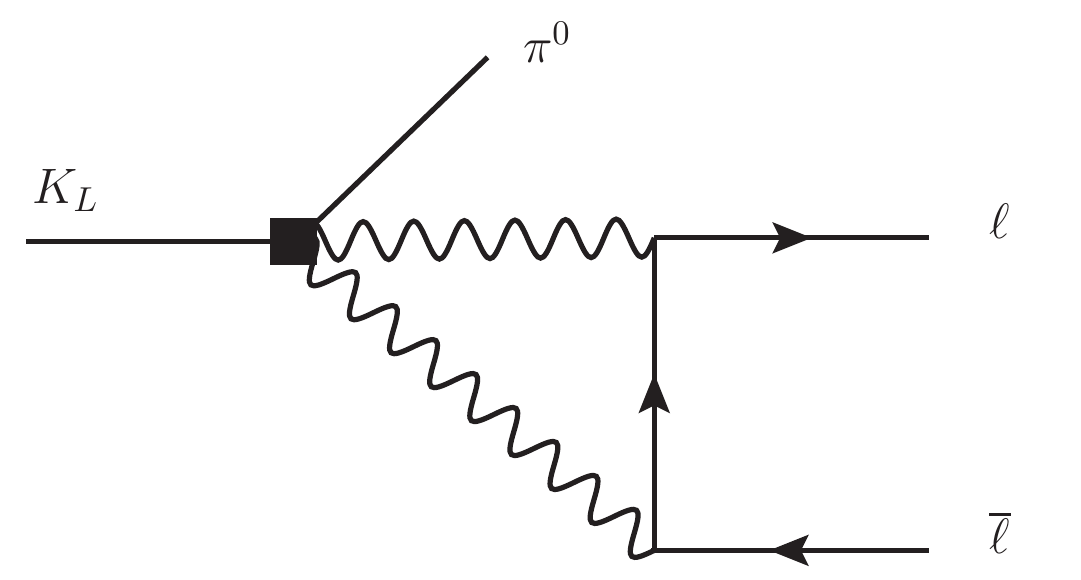}
\caption{\label{fig:05} CP-conserving contribution to $K_L \rightarrow \pi^0 \ell^+ \ell^-$.}
%\end{minipage}
\end{center}
\end{figure}

 
There are three dominating contributions:
\begin{itemize}
\item An indirect CP-violating transition due to the $K^0$-$\overline{K}^0$ mixing, proportional to $V_S(z)$ in Eq.~(\ref{eq:vjz}):
\begin{equation} \label{eq:epsilonvlz}
V_L^{\mathrm{ind}}(z) = \pm \, \varepsilon \, V_S(z)  \, .
\end{equation}
%with $V_S(z)$ given in Eq.~(\ref{eq:vjz}). 
\item A direct CP-violating contribution arising from the four-fermion effective operators $Q_{7V}$ and $Q_{7A}$ given in 
\cite{Cirigliano:2011ny}.
%\begin{equation} \label{eq:7va}
%{\cal L}_{\mathrm{eff}}^{\Delta S = 1} = - \frac{G_F}{\sqrt{2}} V_{ud} V_{us}^* \, [\overline{s} \gamma^{\alpha}(1-\gamma_5)d]  \sum_{\ell=e,%\mu} \left( \, \overline{\ell} \, \gamma_{\alpha} \, [ \, C_{7V}(\mu) + C_{7A}(\mu) \,  \gamma_5 \, ] \,  \ell \right) , %
%\end{equation%}
The contributions to the form factors are proportional to $\mbox{Im} \, \lambda_t$ 
%\begin{eqnarray}\label{eq:klff}
%V_L^{\mathrm{dir}} (z)& = & i \frac{4 \pi}{\sqrt{2} \alpha}  \, y_{7V} \, \mbox{Im} \, \lambda_t \, f_{+}^{K \pi}(z) \, , \nonumber \\
%A(z) & = & i \frac{4 \pi}{\sqrt{2} \alpha}  \, y_{7A} \, \mbox{Im} \, \lambda_t \, f_{+}^{K \pi}(z) \, , \nonumber \\
%P(z) & = & - i \frac{8 \pi}{\sqrt{2} \alpha}  \, y_{7A} \, \mbox{Im} \, \lambda_t \, f_{-}^{K \pi}(z) \, , 
%\end{eqnarray} 
and the  semileptonic form factors of $\langle \pi(p_{\pi}) | \overline{s} \gamma_{\mu} u | K(p_K) \rangle$ \cite{Isidori:2004rb}. 
% $\langle \pi(p_{\pi}) | \overline{s} \gamma_{\mu} u | K(p_K) \rangle = (p_k + p_{\pi})_{\mu} 
%f_{+}^{K \pi}(t) + (p_k - p_{\pi})_{\mu} f_{-}^{K \pi}(t)$ and $t=(p_K -p_{\pi})^2$. 
\item A CP-conserving contribution from $K_L \rightarrow \pi^0 \gamma \gamma \rightarrow \pi^0 \ell^+ \ell^-$ \cite{Heiliger:1992uh}, 
as given by the Feynman diagram in Figure~\ref{fig:05}.
\end{itemize} 
The addition of both CP-violating contributions gives (using the VMD model to eliminate $b_S$):
\begin{eqnarray} \label{eq:cpvl}
10^{12} \times \mathrm{BR}(K_L \rightarrow \pi^0 e^+ e^-)\Big|_{\mathrm{CPV}} &=& 15.7 \, |a_S|^2 \pm 6.2 \,  |a_S| \left( \frac{\mbox{Im} \, \lambda_t}{10^{-4}} 
\right) + 2.4  \left( \frac{\mbox{Im} \, \lambda_t}{10^{-4}} \right)^2 \, , \nonumber \\
10^{12} \times \mathrm{BR}(K_L \rightarrow \pi^0 \mu^+ \mu^-)\Big|_{\mathrm{CPV}} &=& 3.7 \, |a_S|^2 \pm 1.6 \, |a_S| \left( \frac{\mbox{Im} \, \lambda_t}{10^{-4}} 
\right) + 1.0  \left( \frac{\mbox{Im} \, \lambda_t}{10^{-4}} \right)^2 \, .
\end{eqnarray}
It can be seen that for $|a_S| \sim 1$ (as the phenomenological results of Table~\ref{tab:1} suggest) the indirect CP-violating contribution
should be dominating in the electronic case while for the muon both are alike. 
\par 
In Table~\ref{tab:2} we compare the present experimental bounds with both CP-violating and CP-conserving pieces. The dominating contribution to the latter is helicity suppressed ($\propto M_{\ell}$) and, therefore, it is negligible in the electron case. However, although smaller than the CP-violating, the CP-conserving contribution is not tiny in $K_L \rightarrow \pi^0 \mu^+ \mu^-$. It is interesting to note how close to the present experimental bounds are the SM predictions. We are almost there.
\begin{table} 
\centering
\begin{tabular}{|cccc|}
\hline
Process & Theory CP-violating & Theory CP-conserving & Experiment ($90 \%$ C. L.)\\
\hline
$K_L \rightarrow \pi^0 e^+ e^-$ & $3.1 \pm 0.9$ & $\sim 0$ & $< 28$ \cite{AlaviHarati:2003mr} \\
\hline 
$K_L \rightarrow \pi^0 \mu^+ \mu^-$ & $1.4 \pm 0.5$ & $0.52 \pm 0.16$ & $< 38$ \cite{AlaviHarati:2000hs} \\
\hline
\end{tabular}
\caption{\label{tab:2} Comparison between theoretical and experimental results for $\mathrm{BR}(K_L \rightarrow \pi^0 \ell^+ \ell^-)$. 
The numbers correspond to BR$\times 10^{11}$. We assume positive interference between both CP-violating contributions (theoretically 
preferred \cite{Buchalla:2003sj}).}
\end{table}
\section*{Acknowledgments}
This work has been partially supported by MEC (Spain) under grant
FPA2014-53631-C2-1-P], by the Spanish Centro de Excelencia Severo Ochoa 
Programme [SEV-2014-0398] and by the Generalitat Valenciana [PrometeoII/2013/007].
 
\section*{References}
%\bibliography{kaons}
%\providecommand{\newblock}{}

\end{document}